\documentclass[amsmath,amssymb,twocolumn]{revtex4-1}
\usepackage{graphicx}
\usepackage{hyperref}
\usepackage{url}
\begin{document}
\title{Subtraction of test mass angular noise in the LISA Technology Package interferometer}
\author{Felipe Guzm\'an Cervantes}
\email[Corresponding author:]{felipe.guzman@aei.mpg.de}
\author{Frank Steier}
\author{Gudrun Wanner}
\author{Gerhard Heinzel}
\author{Karsten Danzmann}
\affiliation{Albert-Einstein-Institut Hannover (Max-Planck-Institut f\"ur Gravitationsphysik, and Leibniz Universit\"at Hannover) Callinstra\ss e 38, 30167 Hannover, Germany}
%
\begin{abstract}
We present recent sensitivity measurements of the LISA Technology Package interferometer with articulated mirrors as test masses, actuated by piezo-electric transducers. The required longitudinal displacement resolution of $9\,\mathrm{pm}/\sqrt{\mathrm{Hz}}$ above 3\,mHz has been demonstrated with an angular noise that corresponds to the expected in on-orbit operation. The excess noise contribution of this test mass jitter onto the sensitive displacement readout was completely subtracted by fitting the angular interferometric data streams to the longitudinal displacement measurement. Thus, this cross-coupling constitutes no limitation to the required performance of the LISA Technology Package interferometry.
\end{abstract}
\maketitle
\def\registered{{\ooalign{\hfil\raise .00ex\hbox{\tiny R}\hfil\crcr\mathhexbox20D}}}
\section{Introduction}
\label{intro}
The Laser Interferometer Space Antenna (LISA)\,\cite{vitale-1} is a joint space mission from the European Space Agency (ESA) and the National Aeronautics and Space Administration (NASA), designed as a gravitational wave observatory in the frequency range of $0.1\,\mathrm{mHz}$ to $0.1\,\mathrm{Hz}$. LISA consists of a three spacecraft constellation in a equilateral triangle formation, flying a total of six free-falling test masses that act as end-mirrors of laser interferometers sensitive to position fluctuations $\Delta L$ better than 40\,$\mathrm{pm}/\sqrt{\mathrm{Hz}}$ over the interspacecraft separation $L$ of 5 million kilometers. These fluctuations in the separation between two test masses are caused by the space-time distortion caused by gravitational waves, as well as residual acceleration noise. The corresponding strain sensitivity ${\Delta L}/L$ of LISA is of the order of 10\,$^{-21}/\sqrt{\mathrm{Hz}}$.
LISA requires highly challenging technology that is under development and cannot be tested on Earth. To this end, ESA will launch the technology demonstration mission LISA Pathfinder (LPF), which consists of a single satellite carrying two payloads: the LISA Technology Package (LTP) provided by ESA, and the Disturbance Reduction System (DRS) from NASA. LTP\,\cite{anza} is a set of experiments designed to test core technology essential for LISA, such as:
\begin{enumerate}
\item free-fall motion of a test mass with acceleration noise lower than $3\times10^{-14}\,\mathrm{m\,s}^{-2}/\sqrt{\mathrm{Hz}}$ at 1\,mHz,
\item high-precision laser interferometry with a free-falling mirror (LTP test mass) with displacement sensitivity better than $9\times10^{-12}\,\mathrm{m}/\sqrt{\mathrm{Hz}}$ between 3\,mHz and 30\,mHz over a wide dynamic range (several microns),
\item satellite position correction via micronewton thrusters to assure a closed drag-free test mass displacement control loop.
\item assess reliability and lifetime of components in space, such as optics, and lasers, among others.
\end{enumerate}
The main concept of LTP is to shorten one $5\times10^{9}\,\mathrm{m}$ LISA interferometer arm to a distance of about 30\,cm. A laser interferometer is located between the two LTP test masses and measures fluctuations in their separation with a resolution better than $9\,\mathrm{pm}/\sqrt{\mathrm{Hz}}$, as well as their angular orientation with a sensitivity of $10\,\mathrm{nrad}/\sqrt{\mathrm{Hz}}$. The LTP test mass (TM) is a reflecting cube made of a platinum-gold (Pt-Au) alloy and resides in a electrode housing (EH). The electrodes at the EH internal sides and the corresponding faces of the cubic TM form a capacitance that can be measured to obtain the TM position. It is also possible to actuate the TM position by applying an electric field on the TM through the electrodes. The Drag-Free Attitude and Control System (DFACS)\,\cite{fichter-1} uses the optical metrology output and the capacitive sensing as error signals to control a drag-free motion of the TM. The micronewton thrusters are the actuators on the satellite position to close the DFACS control loop. Due to limited gain in the DFACS control loop, the test masses have residual angular noise with respect to the spacecraft that couples into the longitudinal interferometric measurement, thus affecting the performance of the optical metrology system. This article presents investigations conducted on this effect. As free-falling test masses we used articulated mirrors of the interferometer with 3-axes piezo-electric transducers (PZT), and the engineering model of the LTP optical bench was used as optical metrology instrument.
\section{LISA technology package interferometry}
\label{oms}
\begin{figure*}[ht]
\resizebox{0.9\textwidth}{!}{%
\includegraphics{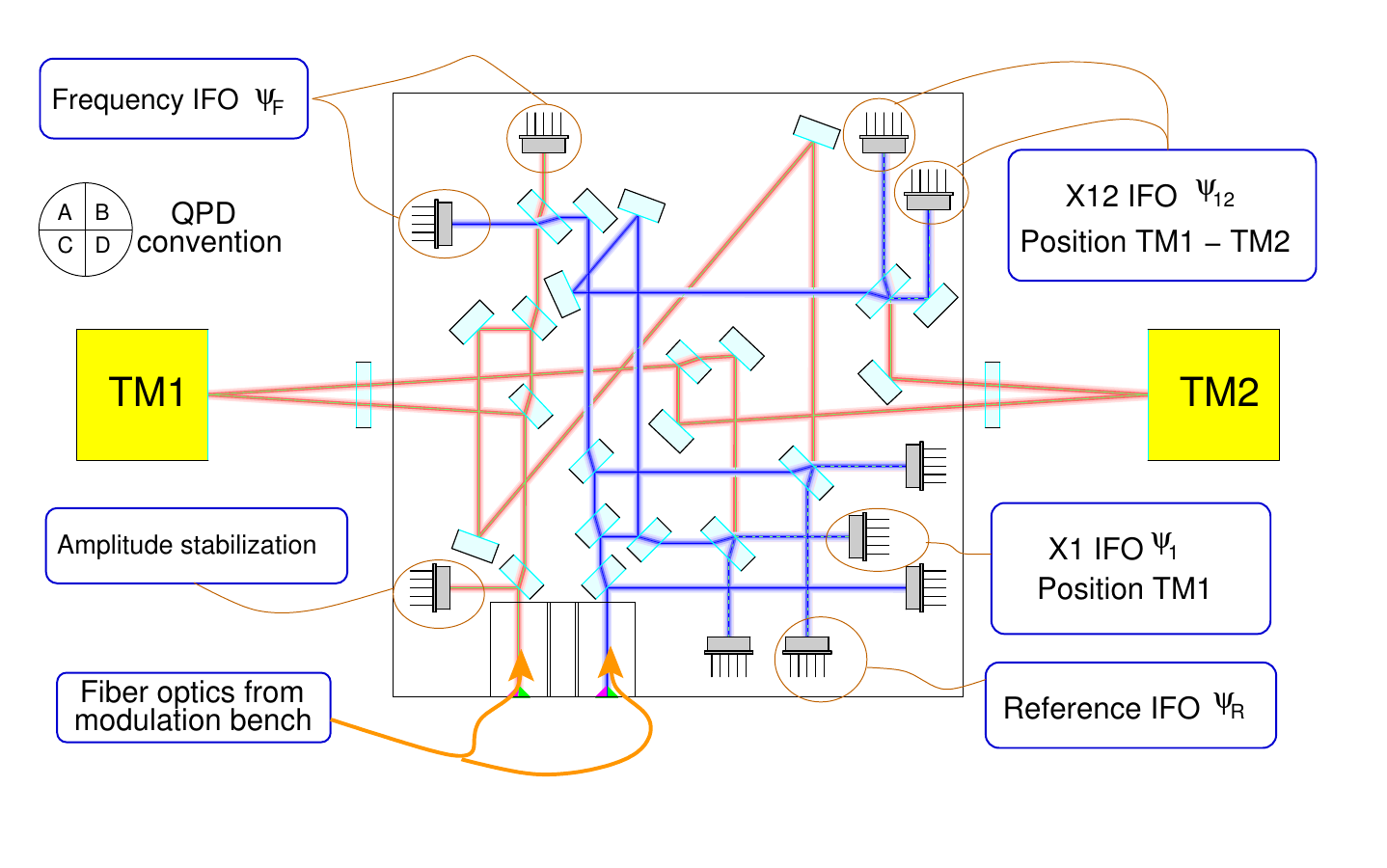}
}
\caption{\label{ifos}Optical model of the LTP optical bench engineering model. Note the interference points of the four interferometers.}
\end{figure*}

\begin{figure}[ht]
\resizebox{0.5\textwidth}{!}{
\includegraphics{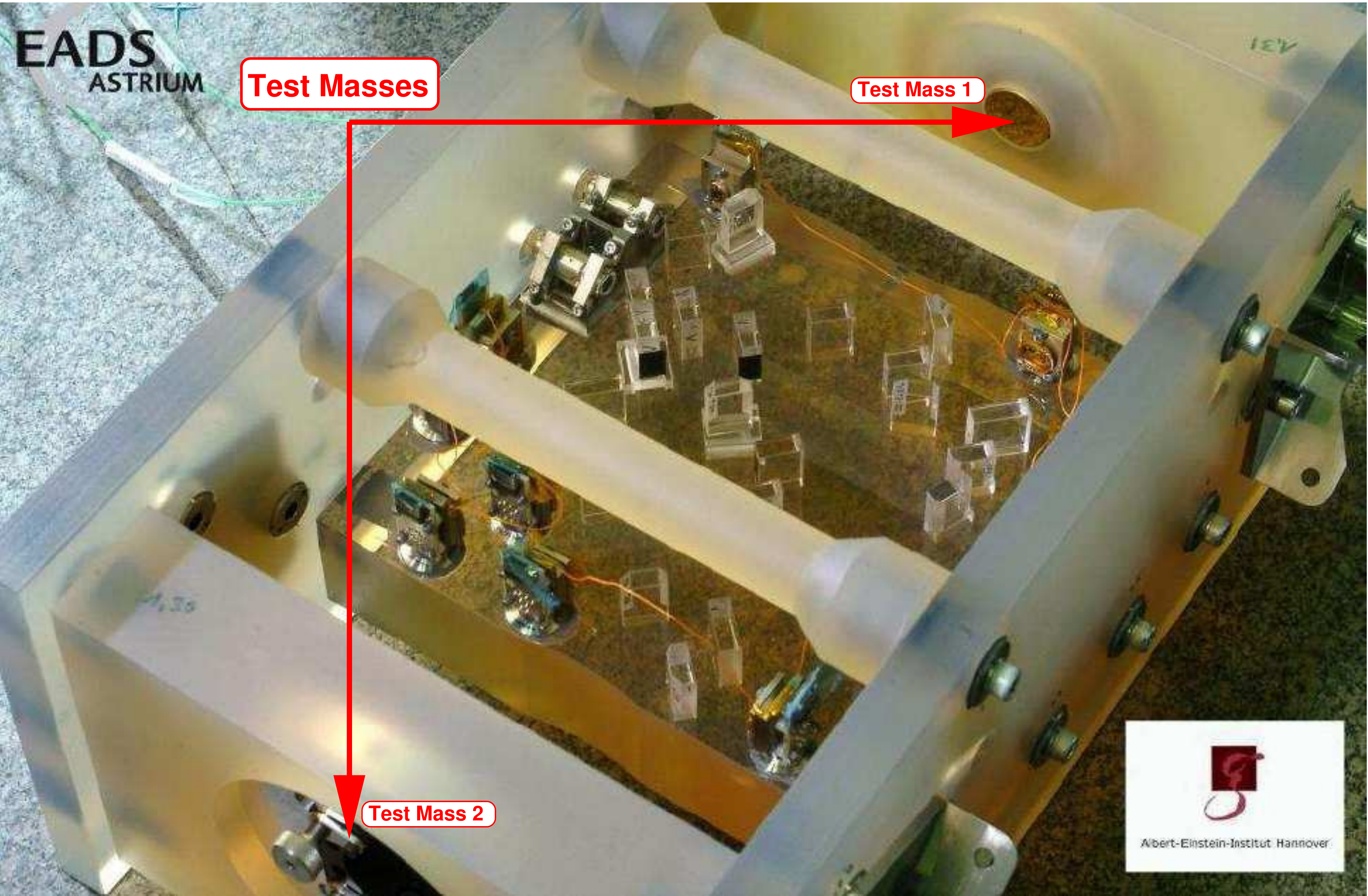}
}
\caption{\label{em-obi}Space-qualified engineering model of the LTP optical bench. Note the expected location of the LTP test masses and the mounting of the test mirrors that simulate them.}
\end{figure}

\begin{figure}[ht]
\resizebox{0.5\textwidth}{!}{
\includegraphics{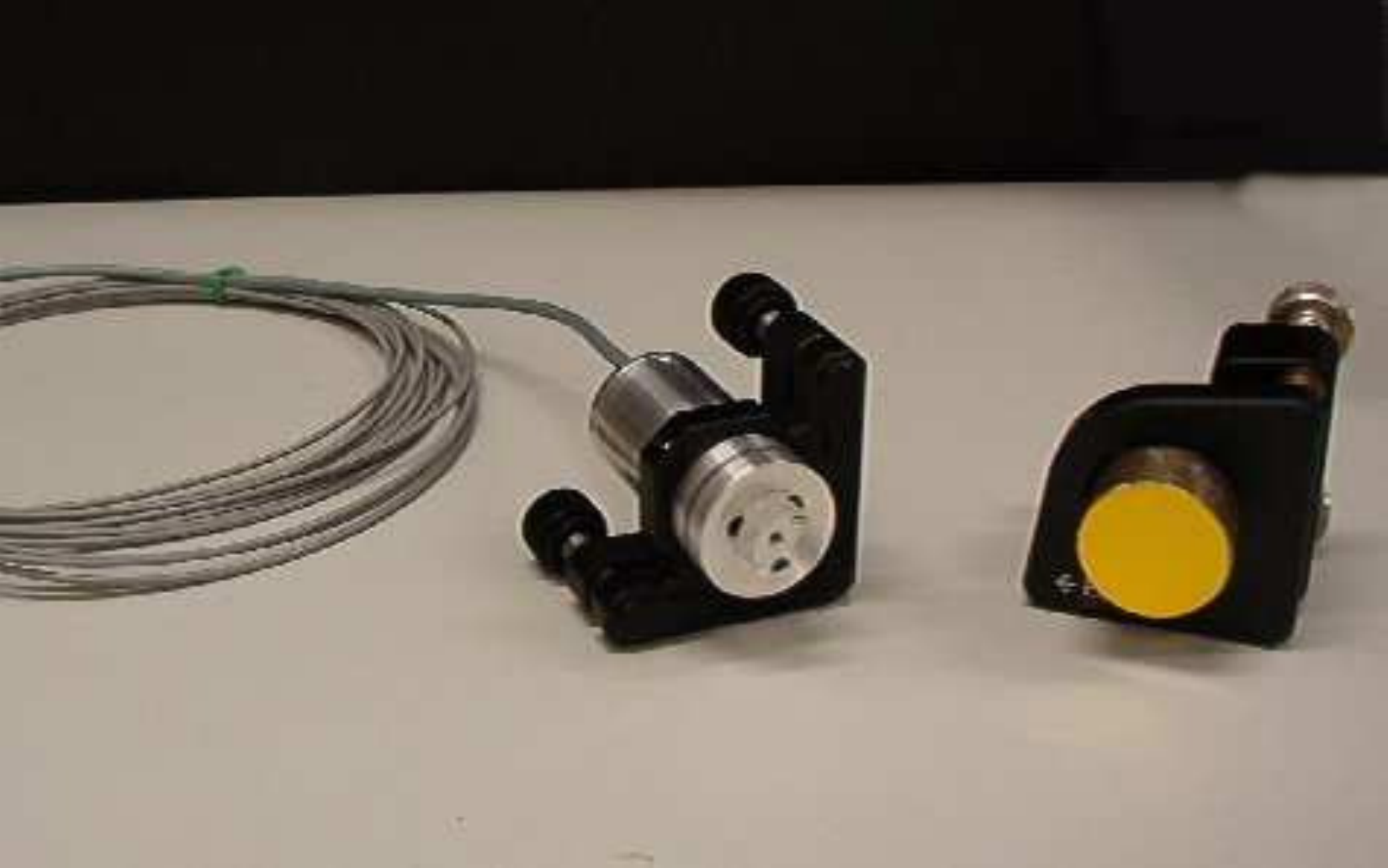}
}
\caption{\label{pztmirror}Comparison between PZT actuated mirrors (left) and static test mirrors (right).}
\end{figure}

\begin{figure}[ht]
\resizebox{0.5\textwidth}{!}{\rotatebox{-90}{
\includegraphics{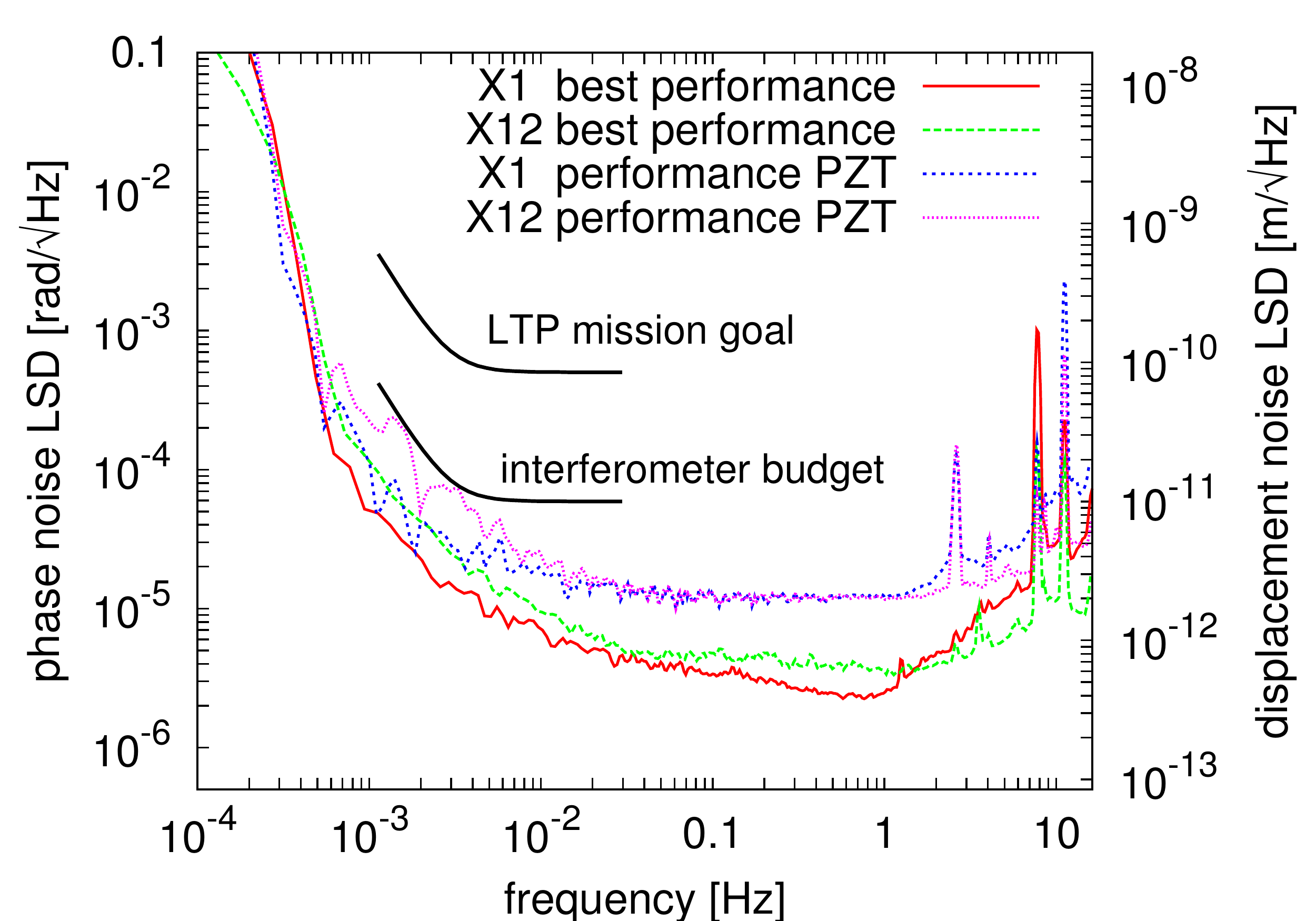}
}}
\caption{\label{pzt-perf}Sensitivity spectra of longitudinal phase measurements performed with static test mirrors and with PZT actuated mirrors.}
\end{figure}

\begin{figure}[ht]
\resizebox{0.5\textwidth}{!}{
\includegraphics{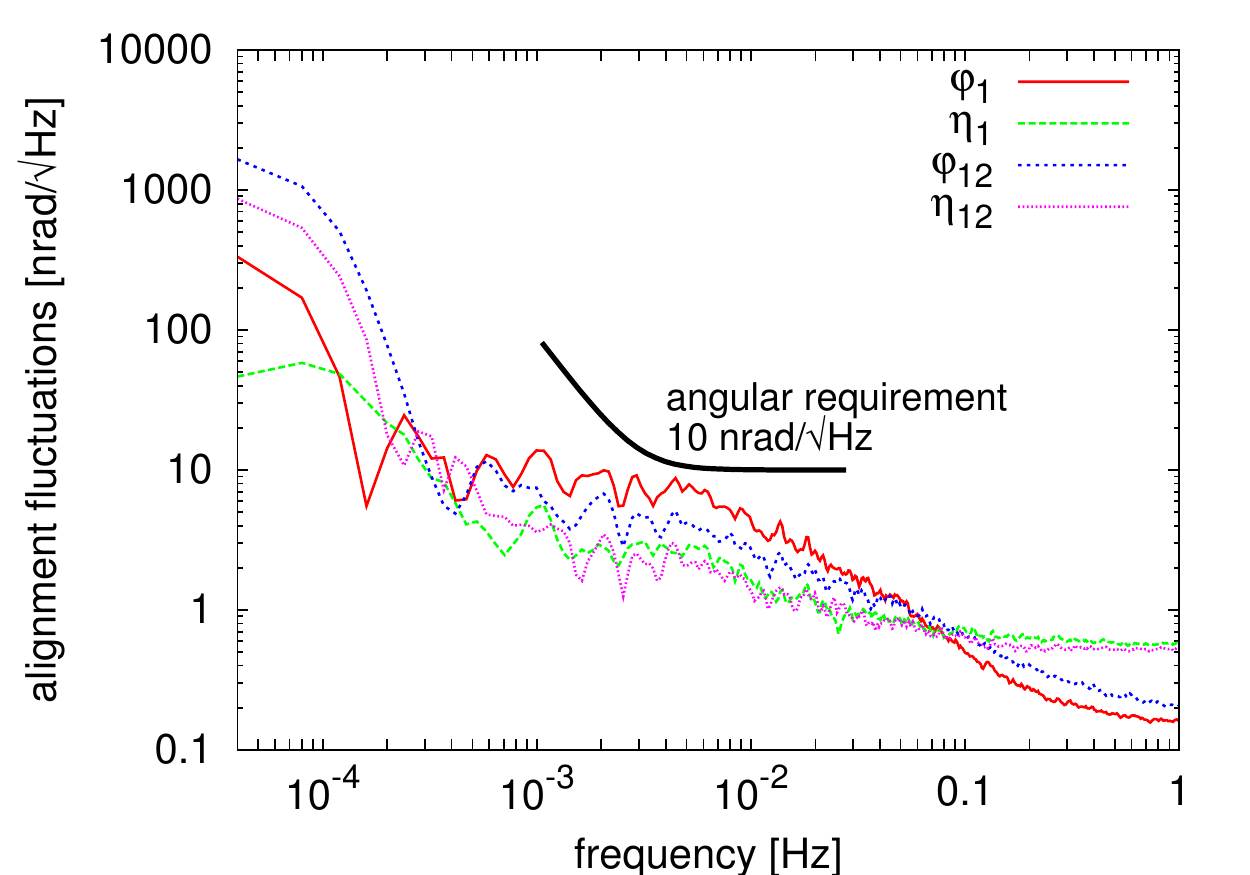}
}
\caption{\label{ang-perf}Sensitivity spectra of angular measurements with PZT actuated mirrors.}
\end{figure}

The LTP interferometer\,\cite{ghh1} is divided into two parts:
\begin{itemize}
\item a ultra-stable optical bench, consisting of a set of four non-polarizing heterodyne Mach-Zehnder interferometers, whose fused-silica optical components are bonded onto a Zerodur$^{\begin{scriptsize}\registered\end{scriptsize}}$ baseplate\,\cite{killow}, what provides high thermal and mechanical stability, and
\item a comparably unstable modulation bench containing the laser source and two acousto-optic modulators that provide two slightly frequency shifted laser beams and respective fiber coupling to transfer the two modulated beams.
\end{itemize}
Figure~\ref{ifos} outlines the optical paths of these four interferometers, which can be described as follows:
\begin{itemize}
\item The X12 interferometer (IFO) measures the fluctuations in the separation between the two drag-free test masses TM\,1\,--\,TM\,2. Beam 1 and Beam 2 overlap at the beam combiner BS10, and the interference signal is obtained from the redundant quadrant photodiodes PD12A and PD12B.
\item The X1 IFO monitors the TM\,1 position fluctuations with respect to the optical bench. Both beams recombine at the beamsplitter BS8, and the interference signal is obtained from the redundant quadrant photodiodes PD1A and PD1B.
\item The reference IFO operates within the ultra-stable optical bench only, detecting disturbances common to all interferometers that couple into the measurement in the unstable part (modulation bench and fiber optics), such that they can be subtracted from X12 and X1. The recombination beamsplitter for this interferometer is BS5 and the readout photodetectors are PDRA and PDRB.
\item The frequency stabilization IFO has an intentionally large optical pathlength difference, in order to sense the laser frequency noise, and its output signal is used to actively stabilize the laser frequency. The two beams overlap at BS7 and the input signal for the laser frequency control loop is obtained from the photodetectors PDFA and PDFB.
\end{itemize}
Figure~\ref{em-obi} is a photograph of the space-qualified engineering model of the LTP optical bench\,\cite{ghh2} and points out the location and mounting of test mirrors that simulate the LTP test masses in our experimental setup. The photocurrents are processed by a dedicated phasemeter\,\cite{ghh3} that performs a single-bin discrete Fourier transform at the heterodyne frequency (difference frequency between the two slightly frequency shifted beams) on field programmable gate array (FPGA) based digital hardware. The phase of each interferometer is computed as the arc tangent between two data streams in orthogonal quadrature. Finally, the longitudinal phase $\mathrm{\phi}$ is obtained from the difference between the reference phase (phase of the reference IFO) and the phase of the measurement interferometers (X1, X12, and frequency stabilization).
The conversion from longitudinal phase $\mathrm{\phi}$ to TM longitudinal displacement $\mathrm{\Psi}$ is
\begin{equation}
\mathrm{\Psi}=\frac{\lambda}{4\pi}\phi\,,
\end{equation}
where $\lambda=1064\,\mathrm{nm}$ is the wavelength of the Nd:YAG laser used. All photodetectors at the output of all four interferometer are quadrant photodiodes (QPD), in order to obtain alignment information and sense angular motion of the test masses by applying a differential wavefront sensing (DWS)\,\cite{morrison-1,morrison-2} technique. The horizontal wavefront tipping $\varphi$ can be calculated from the phase difference between the left (quadrants A and C) and the right side (quadrants B and D) of the QPD. Similarly, the vertical wavefront tipping $\eta$ is obtained from the phase difference between the upper (quadrants A and B) and lower side (quadrants C and D) of the QPD. The main output of the optical metrology is
\begin{itemize}
\item $\mathrm{\Psi}_{1}$: longitudinal position fluctuations of TM\,1 with respect to the optical bench,
\item $\varphi_{1}$: horizontal angular motion of TM\,1 in the X1 IFO,
\item $\eta_{1}$: vertical angular motion of TM\,1 in the X1 IFO,
\item $\mathrm{\Psi}_{12}$: longitudinal distance fluctuations between TM\,1 and TM\,2,
\item $\varphi_{12}$: combination of the horizontal angular motion of TM\,1 and TM\,2 in the X12 IFO, and
\item $\eta_{12}$: combination of the vertical angular motion of TM\,1 and TM\,2 in the X12 IFO.
\end{itemize}
The requirements on the LTP interferometric sensitivity have been met with Zerodur$^{\begin{scriptsize}\registered\end{scriptsize}}$ static mirrors\,\cite{ghh4} shown in Figure~\ref{em-obi}. Diverse noise sources in the system have been studied and corrected and the performance has been experimentally demonstrated on the engineering model of the LTP optical bench\,\cite{vinne-1}. The optical bench is normally operated in a vacuum chamber to reduce the effect of acoustic noise, and thermal and mechanical fluctuations in the optical measurement. In order to investigate the effect of test mass residual angular noise into the longitudinal measurement, the Zerodur$^{\begin{scriptsize}\registered\end{scriptsize}}$ static mirrors were substituted by PZT actuated mirrors, which are shown in Figure~\ref{pztmirror}.
Figure~\ref{pzt-perf} shows the sensitivity spectra reached for the longitudinal TM displacement $\mathrm{\Psi}_{1}$ and $\mathrm{\Psi}_{12}$ measured with Zerodur$^{\begin{scriptsize}\registered\end{scriptsize}}$ static mirrors and with forward biased PZT actuated mirrors in static condition. It can be seen that all sensitivity curves remain below the required $9\,\mathrm{pm}/\sqrt{\mathrm{Hz}}$ in the measurement band (interferometer budget).
Figure~\ref{ang-perf} shows the horizontal ($\varphi_{1,12}$) and vertical ($\eta_{1,12}$) angular resolution spectra achieved with forward biased PZT mirrors in static condition, which is better than the required $10\,\mathrm{nrad}/\sqrt{\mathrm{Hz}}$ TM jitter in the measurement band.
\section{Test mass angular noise characterization}
\label{tmnoise}
\begin{figure}[ht]
\resizebox{0.5\textwidth}{!}{
\includegraphics{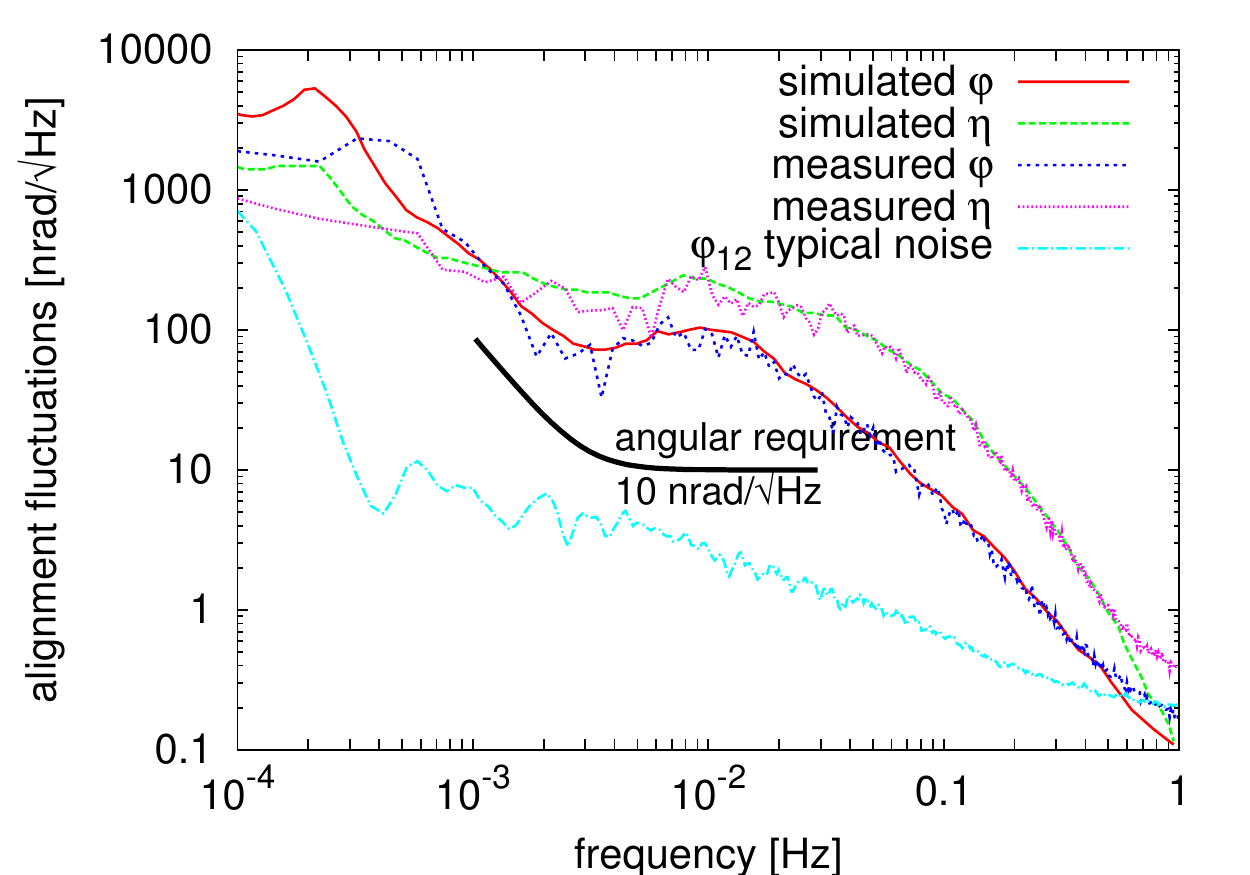}
}
\caption{\label{ang-noise}Expected injected residual TM angular noise for on-orbit operation and interferometric measured angular noise of the test masses.}
\end{figure}

Simulations conducted on the TM dynamics under DFACS control led to spectral predictions of the residual TM angular noise\,\cite{fichter-email}. From this spectral information, we generated a time series that matches this spectral behavior and injected it to the PZT actuated mirrors via a digital analog converter (DAC). Figure~\ref{ang-noise} shows the simulated angular noise spectra ($\varphi_{1,12}$ and $\eta_{1,12}$), and the corresponding TM angular noise spectra read out by the interferometers and computed with a DWS algorithm.
Due to limited accuracy alignment of the laser beam onto the TM center of rotation, the resultant cross-coupling from TM angular noise into the TM longitudinal displacement readout introduces excessive noise into the measurement of TM position fluctuations, thus spoiling the sensitivity of the optical readout. Our aim is to characterize this cross-coupling and to quantitatively obtain the corresponding coupling factors that translate this angular motion of the TM into an apparent longitudinal TM displacement. Once these coupling factors have been estimated, the excessive noise can be subtracted from the main longitudinal data stream.
\subsection{Fit algorithm}
\label{fit}
The longitudinal raw measurements $\mathrm{\Psi}_{1}$ and $\mathrm{\Psi}_{12}$ depend on the TM angular motion $\varphi_{1,12}$ and $\eta_{1,12}$
\begin{eqnarray}
\mathrm{\Psi}_{1} & = & \mathrm{\Psi}_{1}\left( t,\varphi_{1},\eta_{1}\right)\\
\mathrm{\Psi}_{12}& = & \mathrm{\Psi}_{12}\left( t,\varphi_{1},\eta_{1},\varphi_{12},\eta_{12}\right).
\end{eqnarray}
As it can be seen in Figure~\ref{ang-noise}, the signal-to-noise ratio of the TM angular motion is much higher within the LTP observation band ($3\,\mathrm{mHz}\,-\,30\,\mathrm{mHz}$) than at higher frequencies. Under normal laboratory conditions, the effect of fast electronic and mechanical noise in the higher frequency band (approximately above $100\,\mathrm{mHz}$), as well as long-term thermal drifts at frequencies below 1\,mHz dominate the time evolution and behavior of the longitudinal and angular interferometric signals. Hence, the information of the TM angular noise vanishes in the noise level of the measured time series. In order to overcome this limitation, we decided to band-pass filter each longitudinal ($\mathrm{\Psi}_{1}^{\rm bp}$ and $\mathrm{\Psi}_{12}^{\rm bp}$) and angular ($\varphi_{1}^{\rm bp},\eta_{1}^{\rm bp}$ and $\varphi_{12}^{\rm bp},\eta_{12}^{\rm bp}$) time series within the LTP observation band. This way, we can precisely characterize the cross-coupling process of the TM angular motion into the longitudinal interferometric readout. The dependency of the filtered longitudinal measurements $\mathrm{\Psi}^{\rm bp}$ with respect to the filtered angular signals $\varphi^{\rm bp},\eta^{\rm bp}$ can be rephrased as
\begin{eqnarray}
\mathrm{\Psi}_{1}^{\rm bp} & = & \mathrm{\Psi}_{1}\left(\varphi_{1}^{\rm bp},\eta_{1}^{\rm bp}\right) \\
\mathrm{\Psi}_{12}^{\rm bp}& = & \mathrm{\Psi}_{12}\left(\varphi_{1}^{\rm bp},\eta_{1}^{\rm bp},\varphi_{12}^{\rm bp},\eta_{12}^{\rm bp}\right).
\end{eqnarray}
Detailed optical simulations have shown a nonlinear coupling mechanism of parabolic type, but have also shown that for the noise levels occuring in our experiment, a linear model is sufficient. A general model for this approach can be described by the following linear system of equations:
\begin{equation}
\mathbf{\Theta}\cdot\mathbf{\kappa}=\mathbf{\Psi},
\end{equation}
where $\mathbf{\Theta}$ is the design matrix for our fitting problem (angular data $\varphi^{\rm bp},\eta^{\rm bp}$), $\kappa$ is a vector containing the coupling factors we are looking for, and $\mathbf{\Psi}$ is a vector correspondig to the time series of our target function (longitudinal TM data $\mathrm{\Psi}^{\rm bp}$). The dimensions of $\mathbf{\Theta}$ are $N\times m$, where $N$ is the length of the time series and $m$ is the number of input time series to be used; in the case of the X1 IFO $m\,=\,2$ ($\varphi_{1}^{\rm bp},\eta_{1}^{\rm bp}$), and for the X12 IFO $m\,=\,4$ ($\varphi_{1}^{\rm bp},\eta_{1}^{\rm bp},\varphi_{12}^{\rm bp},\eta_{12}^{\rm bp}$). $\kappa$ is a vector with dimensions $m\times 1$, and $\mathbf{\Psi}$ is a vector with dimensions $N\times 1$.\\
In our specific case, we have the following system of equations for the X1 IFO:
\begin{equation}
\mathbf{\Psi}^{1}_{N\times 1}  =  \mathbf{\Theta}^{1}_{N\times 2}\cdot\mathbf{\kappa}^{1}_{2\times 1},
\end{equation}
with
\begin{equation}
\mathbf{\Theta}^{1} = \left( \begin{array}{cc} \varphi_{1}^{\rm bp} & \eta_{1}^{\rm bp} \end{array} \right)\,\,\mathrm{and}\,\, \mathbf{\kappa}^{1} = \left( \begin{array}{cc} \mathrm{\kappa}^{1}_{0} \\ \mathrm{\kappa}^{1}_{1} \end{array} \right).
\end{equation}
The system of equations for the X12 IFO can be expressed as:
\begin{equation}
\mathbf{\Psi}^{12}_{N\times 1}  = \mathbf{\Theta}^{12}_{N\times 4}\cdot\mathbf{\kappa}^{12}_{4\times 1},
\end{equation}
with
\begin{equation}
\mathbf{\Theta}^{12} = \left( \begin{array}{cccc} \varphi_{1}^{\rm bp} & \eta_{1}^{\rm bp} & \varphi_{12}^{\rm bp} & \eta_{12}^{\rm bp} \end{array} \right)\,\,\mathrm{and}\,\,\mathbf{\kappa}^{12} = \left( \begin{array}{cccc} \mathrm{\kappa}^{12}_{0} \\ \mathrm{\kappa}^{12}_{1} \\ \mathrm{\kappa}^{12}_{2} \\ \mathrm{\kappa}^{12}_{3} \end{array} \right).
\end{equation}
The fit can be performed by a general linear least squares algorithm. This linear system of equations can be solved by applying different algorithms such as the Cholesky decomposition, the use of normal equations, or the singular value decomposition, among others. The proper selection of the solving method usually depends on the topology of the design matrix $\mathbf{\Theta}$. This way, we obtain the set of coupling coefficients $\mathbf{\kappa}^{1}$ and $\mathbf{\kappa}^{12}$ of the TM angular noise into the longitudinal TM displacement readout.
\subsection{Test mass angular noise subtraction}
\label{noisesub}
\begin{figure}[ht]
\resizebox{0.5\textwidth}{!}{
\includegraphics{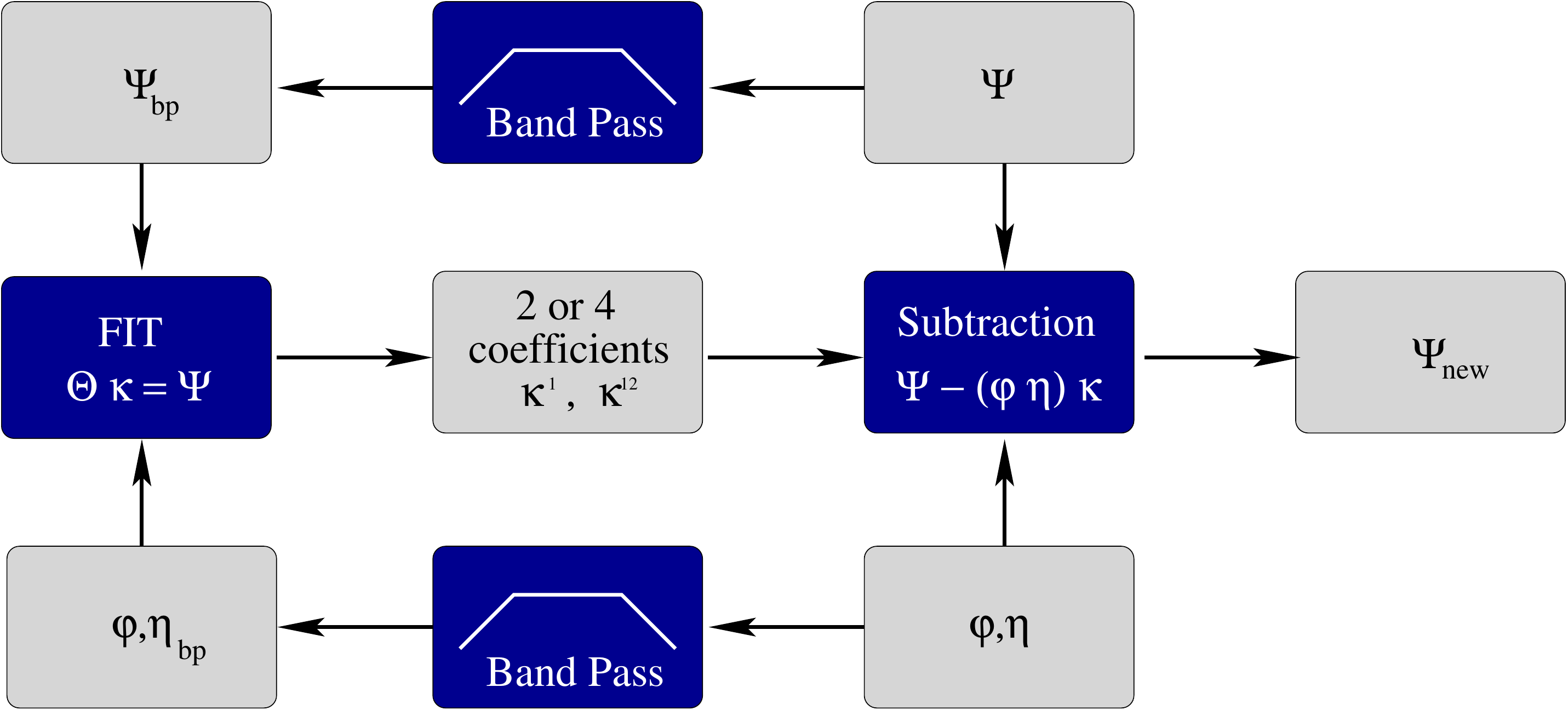}
}
\caption{\label{algo}Flow diagram of the procedure to subtract the TM angular noise from the longitudinal phase data stream.}
\end{figure}

\begin{figure}[ht]
\resizebox{0.5\textwidth}{!}{\rotatebox{-90}{
\includegraphics{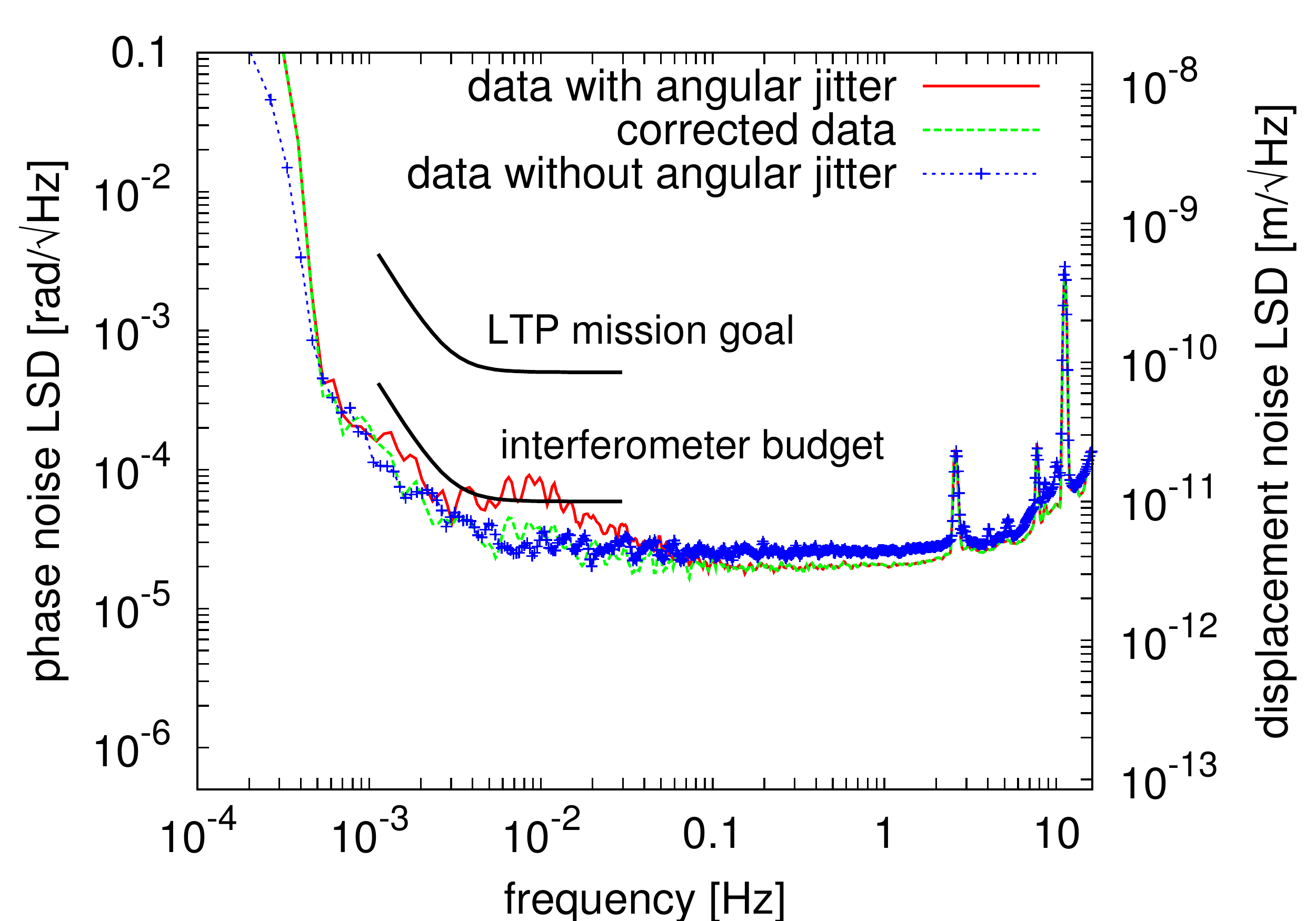}
}}
\caption{\label{corspec}Solid curve: sensitivity of the longitudinal phase readout when injecting TM angular noise. Dashed curve: sensitivity of the corrected longitudinal phase after the angular noise subtraction. Dashed curve with crosses: independent reference measurement with no injected angular noise.}
\end{figure}

The band-pass filtered data $\mathrm{\Psi}^{\rm bp}$ and $\varphi^{\rm bp},\eta^{\rm bp}$ are utilized to obtain the coupling coefficients $\mathbf{\kappa}$. For example, typical fitted values for $\mathbf{\kappa}^{1}$ are
\begin{equation}
\mathrm{\kappa}^{1}\,\left[\mathrm{m}/\mathrm{rad}\right]  = \left( \begin{array}{cc} -4.38\times10^{-6} \\ -2.19\times10^{-5} \end{array} \right).
\end{equation}
Once we have estimated them, it is possible to subtract the TM angular noise from the original (unfiltered) measured longitudinal TM data as follows:
\begin{eqnarray}
\mathrm{\Psi}^{1}_{\rm new}   & = & \mathrm{\Psi}_{1}  \, - \, \left( \begin{array}{cc} \varphi_{1}  & \eta_{1} \end{array} \right) \cdot \mathbf{\kappa}^{1},\,\,\mathrm{and}\\
\mathrm{\Psi}^{12}_{\rm new}  & = & \mathrm{\Psi}_{12} \, - \, \left( \begin{array}{cc} \varphi_{12} & \eta_{12} \end{array} \right) \cdot\mathbf{\kappa}^{12}.
\end{eqnarray}
The entire procedure to subtract the TM angular noise from the longitudinal interferometric signal is outlined by Figure~\ref{algo}.
Figure~\ref{corspec} presents the results obtained from this subtraction.
The solid curve is the sensitivity reached by the longitudinal phase readout $\mathrm{\Psi}$ when introducing TM angular noise (note that it exceeds the required noise budget). The dashed curve for $\mathrm{\Psi_{\rm new}}$ is the sensitivity achieved after subtracting the fitted angular noise to the data of the solid curve. The dashed curve with crosses is the sensitivity obtained from an independent measurement where no angular noise was injected to the test masses (PZT actuated mirrors). In general, noise subtraction procedures have to be performed very carefully, since there is a non-vanishing probability to corrupt the data. In our case, the longitudinal and angular degrees of freedom (DOF) are sufficiently orthogonal, such that the cross-coupling between them can be very well quantitatively characterized. The linear transformation between these two reference systems has been experimentally measured and can be expressed, for example for TM\,1, as
\begin{equation}
\left( \begin{array}{c} \mathrm{\Psi}\,\left[\mathrm{m}\right] \\ \varphi\,\left[\mathrm{rad}\right] \\ \eta\,\left[\mathrm{rad}\right] \end{array} \right)^{\mathrm{OB}}  = \Lambda \cdot \left( \begin{array}{c} \mathrm{\Psi}\,\left[\mathrm{m}\right] \\ \varphi\,\left[\mathrm{rad}\right] \\ \eta\,\left[\mathrm{rad}\right] \end{array} \right)^{\mathrm{TM}}\,,
\end{equation}
where
\begin{eqnarray}
\Lambda & = & \left( \begin{array}{ccc} \frac{\partial\mathrm{\Psi}^{OB}}{\partial\mathrm{\Psi}^{TM}} & \frac{\partial\mathrm{\Psi}^{OB}}{\partial\varphi^{TM}} & \frac{\partial\mathrm{\Psi}^{OB}}{\partial\eta^{TM}} \\\\ \frac{\partial\varphi^{OB}}{\partial\mathrm{\Psi}^{TM}} & \frac{\partial\varphi^{OB}}{\partial\varphi^{TM}} & \frac{\partial\varphi^{OB}}{\partial\eta^{TM}} \\\\ \frac{\partial\eta^{OB}}{\partial\mathrm{\Psi}^{TM}} & \frac{\partial\eta^{OB}}{\partial\varphi^{TM}} & \frac{\partial\eta^{OB}}{\partial\eta^{TM}} \end{array} \right)\\
& = & \left( \begin{array}{ccc} 1 & -4.38\times10^{-6} & -2.19\times10^{-5} \\ 0.6 & 1 & 5.2\times10^{-4} \\ 0.4 & 7.0\times10^{-3} & 1 \end{array} \right).
\end{eqnarray}
An example of a problematic situation where noise subtraction would be expected to corrupt signal is if $\mathrm{\Psi}$ couples into $\varphi$, and $\varphi$ back again into $\mathrm{\Psi}$, with factors such that,
\begin{equation}
\frac{\partial\mathrm{\Psi}}{\partial\varphi}\cdot\frac{\partial\varphi}{\partial\mathrm{\Psi}}\approx1.
\end{equation}
In our case, however, this product is of the order of $10^{-6}$ such that no real signal $\mathrm{\Psi}$ is subtracted. A more detailed analysis is currently under investigation. Typical values for the TM motion are of the order of
\begin{equation}
\left( \begin{array}{c} \mathrm{\Psi} \\ \varphi \\ \eta \end{array} \right)^{\mathrm{TM}}  = \left( \begin{array}{c} 9\times10^{-12}\,\mathrm{m}_{\mathrm{rms}} \\ 1\times10^{-7}\,\mathrm{rad}_{\mathrm{rms}} \\ 3\times10^{-7}\,\mathrm{rad}_{\mathrm{rms}} \end{array} \right).
\end{equation}
As it can be seen in Figure~\ref{corspec}, the corrected data reaches the same level of the reference measurement where no angular noise was applied, which indicates that the complete cross-coupling effect from the TM angular noise into the longitudinal measurement was fitted and extracted without corrupting the data. Hence, the residual TM jitter due to the limited DFACS gain is not a limiting factor to the sensitivity of the interferometric longitudinal TM position measurement in LTP.
\section{Conclusions}
We have presented current sensitivity curves of the LTP interferometry, measured at the engineering model of the optical bench. We have also demonstrated that it is possible to reach picometer resolution in the optical readout at a few mHz even with test masses that are articulated by forward biased piezo-electric transducers. This is an important conclusion for LISA, where the test masses will have comparable angular jitter. Furthermore, we performed experimental investigations on the noise contribution of residual test mass angular noise to the longitudinal test mass displacement, concluding that this cross-coupling process can be fully characterized and completely extracted from the longitudinal measurement data stream. We obtained coupling factors for the angular fluctuations, by fitting the measured angular data series to the longitudinal data with a linear least squares algorithm, using only the expected noise but no additional calibration signal. This method can also be used in other applications to characterize noise sources of different kind of systems.
\end{document}